\documentclass[aps,prl,twocolumn,amsmath,amssymb,nofootinbib,showpacs,superscriptaddress]{revtex4-1}
\usepackage[pdfview=FitV,pdfstartview=FitV]{hyperref}
\usepackage{amsmath,amssymb,amsfonts,empheq,bm}
\usepackage{graphicx}
\usepackage{epsfig}
\usepackage{color}

\begin{document}
\title{Duality in power-law localization in disordered one-dimensional systems}
\author{X. Deng}
\affiliation{Institut f\"ur Theoretische Physik, Leibniz Universit\"at Hannover, Appelstr. 2, 30167 Hannover, Germany}

\author{V. E. Kravtsov}
\affiliation{Abdus Salam International Center for Theoretical Physics, Strada Costiera 11, 34151 Trieste, Italy}
\affiliation{ L. D. Landau Institute for Theoretical Physics, Chernogolovka, 142432, Moscow region, Russia}

\author{G. V. Shlyapnikov}
\affiliation{LPTMS, CNRS, Universite Paris Sud, Universite Paris-Saclay, Orsay 91405, France}
\affiliation{SPEC, CEA, CNRS, Universite Paris-Saclay, CEA Saclay, Gif sur Yvette 91191, France}
\affiliation{Russian Quantum Center, Skolkovo, Moscow 143025, Russia}
\affiliation{\mbox{Van der Waals-Zeeman Institute, University of Amsterdam, Science Park 904, 1098 XH Amsterdam, The Netherlands}}
\affiliation{Wuhan Institute of Physics and Mathematics, Chinese Academy of Sciences, 430071 Wuhan, China}

\author{L. Santos}
\affiliation{Institut f\"ur Theoretische Physik, Leibniz Universit\"at Hannover, Appelstr. 2, 30167 Hannover, Germany}

\date{\today}

\begin{abstract}
The transport of excitations between pinned particles in many physical systems may be mapped to single-particle models with power-law hopping, $1/r^a$.
For randomly spaced particles, these models present an effective peculiar disorder that leads to surprising localization properties. We show that in one-dimensional systems almost
all eigenstates (except for a few states close to the ground state)  are power-law localized for any value of $a>0$. Moreover, we show that our model is an example of a new universality class of models with power-law hopping,
characterized by a duality between systems with long-range hops~($a<1$) and short-range hops~($a>1$)  in which the wave function amplitude falls off algebraically with the same power $\gamma$ from the localization center.
\end{abstract}

\maketitle


Long-range interactions are crucial in many disordered systems. In particular, dipole-induced transport of
excitations, with a hopping amplitude decaying with interparticle distance $r$ as $1/r^3$, plays a key role in disparate scenarios,
including magnetic atoms~\cite{DePaz2013}, polar molecules~\cite{Yan2013}, Rydberg atoms~\cite{Saffman2010}, nitrogen-vacancy centers in diamonds~\cite{Waldherr2014}, and
nuclear spins in solid-state systems~\cite{Alvarez2015}.
Interestingly, systems with tunable power-law interactions, $1/r^a$, have been recently proposed. In the presence of laser-driven coupling, trapped ions may realize
power-law-decaying spin interactions with tunable $0\!<\!a\!<\!3$~\cite{Richerme2014,Jurcevic2014}. Spin models with an arbitrary power law interaction
can be also engineered between atoms trapped in a photonic crystal waveguide~\cite{Hung2016}.
These tunable systems can be realized in low dimensions, opening intriguing questions on transport
with truly long-range interactions, i.e. for $a<d$, with $d$ the dimensionality.

Most of the previous works considered excitation transport amongst regularly spaced particles, e.g by pinning them in an uniformly filled lattice with on-site disorder.
In addition, long-range hopping
has been considered either anisotropic with zero angular average (e.g. dipole hopping~\cite{Levitov1990, AltshAleiEfet2011})
or random with zero ensemble average
~\cite{MirlinFyod1996}.
Under such conditions  the interplay of on-site disorder and
long-range hops may result in localization, critical behavior, or fully extended states. In his seminal work, Levitov suggested that these cases occur
for $a\!>\!d$, $a\!=\!d$ and $a\!<\!d$, respectively~\cite{Levitov1990}.
Later studies for $d\!=\!1$ using a supersymmetric non-linear sigma-model approach confirmed this suggestion~\cite{MirlinFyod1996}. Intense efforts have been devoted to
the critical Power-Law Banded Random Matrix (PLBRM) ensemble~($d\!=\!a\!=\!1$)~\cite{KrMut1997,KrYev2003,YevOss2007,Cuevas2007,Chalk2010,Evers2008,RMTbook}.
Dipolar excitations in a 2D lattice with on-site disorder~($d\!=\!a\!=\!2$) exhibit a purely critical behavior only for the time-reversal-invariant case and
may have a "metal-insulator" transition if the invariance is broken~\cite{AltshAleiEfet2011}.

Notwithstanding these details, the common wisdom until recently was that for non-interacting particles with on-site disorder
long-range hopping is a delocalizing factor that destroys localization for $a\leq d$. Only few recent works have stood out of this paradigm.
Surprisingly, all states appeared to be localized
for the exotic case of linear hopping~($a=-1$) between randomly placed points in a 1D system~\cite{Bogomol2003}.
A second example is provided by a {\it simplex}~($a=0$), where all excited states were found to be localized~\cite{Ossipov2012}.
This peculiar behavior was linked in Ref.~\cite{Ossipov2012}  to the macroscopic degeneracy of the perturbation matrix.
The role of degeneracy in this model was recently further emphasized and extended to $a>0$ in Ref.~\cite{Celardo2016}, where it was argued that
the long-range hopping has no effect on the system dynamics as long as the width of the
disorder-broadened $(N-1)$-fold degenerate level~(with $N$ the number of sites) is smaller than the gap between the ground and excited states for $a=0$. Since this gap is proportional to
$N^{1-a}$ and the bandwidth is $N$-independent, the long-range hopping is "shielded" in the thermodynamic limit for all $0<a<1$ and the localization of excited states is preserved.


In this Letter we consider the physically relevant case of excitation transport due to power-law hops amongst randomly pinned particles.
We call these models {\it power-law Euclidean}~(PLE) models in connection with the Euclidean random matrices with
no diagonal elements~\cite{MezarParisiZee1999}.
Disorder in PLE models is {\it purely off-diagonal } given by the Poisson distribution of the particle positions.
However, it is crucially different from the off-diagonal disorder in nearest-neighbor
hopping models \cite{Dyson1953, Lifshitz1964},
since long-range hops break down the chiral~(sublattice) symmetry~\cite{Brouwer2000}  of the nearest-neighbor hopping case.
It also differs from the off-diagonal disorder of Ref.~\cite{MirlinFyod1996}
because the angular or ensemble average of the hopping amplitude is not zero.

We focus on the 1D case, showing that PLE models possess features radically different from those with random-sign long-range hopping~\cite{MirlinFyod1996}.
Similar to the Anderson model on a lattice with diagonal disorder and non-random long-range hopping~\cite{Celardo2016}, almost all states~(except for a few states close to the lower edge of the spectrum, see Suppl. Material)
are localized  for {\it any} $a>0$. However, in contrast to the conventional Anderson model with exponential localization,
in the PLE model localization is {\it algebraic} with a power-law decrease of wave functions,
${\rm exp}\left(\overline{\ln|\psi(x)|^{2}}\right)\sim |x-x'|^{-\gamma}$,
where $\gamma$ depends on $a$.
This remains valid for the model of Ref.~\cite{Celardo2016}, even for arbitrarily small but finite $a$. As discussed below, power-law localization emerges because
 long-range hops are not fully shielded, but rather become effectively short-range.

\begin{figure}[t]
\begin{center}
\includegraphics[width =0.9\columnwidth]{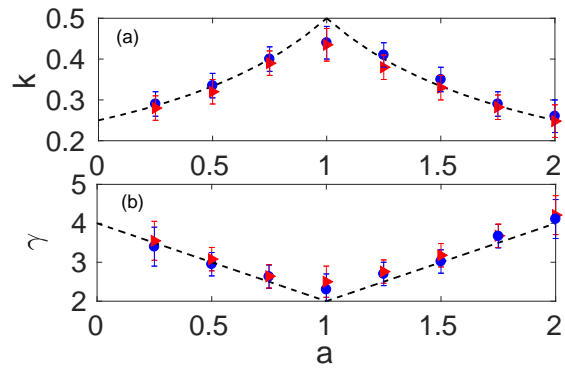}
\caption{Duality  for $0<a<1$ and $1<a<2$ of the localization properties of the PLE model~(blue circles) and the model of Refs.~\cite{Celardo2016,Ossipov2012}
with diagonal disorder and deterministic, sign-definite, long-range hopping~(red triangles): (a) slope $k$ of the MFS extrapolated to $N\to\infty$; (b) exponent $\gamma$ of the algebraic localization of $|\psi|^2$.
Dashed lines correspond to $\gamma=1/k = 2a$ for $1<a<2$ or to $\gamma=4-2a$ for  $0<a<1$.
The error bars mainly stem from the extrapolation to infinite size systems.}
\label{Fig:duality}
\end{center}
\end{figure}

The main result of this paper is a surprising {\it duality}:
\begin{equation}
\gamma(a) = \gamma(2-a),
\label{duali}
\end{equation}
for $0\!<\!a\!<\!1$. We obtain this duality numerically for $0<a<1$~(Fig.~\ref{Fig:duality}), and analytically for the PLE model at finite but small $a$~\cite{footnote-a0}.
Strikingly, this duality is more general than our PLE models.
It holds for the model of Ref.~\cite{Celardo2016},
which presents on-site disorder in a regular lattice with long-range hops.
It also holds for the PLE model with additional on-site disorder.
Moreover, the exponent $\gamma$ depends only on  $a$ but not on other details of the models, including the eigenstate energy, provided that it is in the bulk of the spectrum.
%
%
%
Thus the considered models are not only experimentally relevant but represent a new {\it class} of long-range models where localization is algebraic and the duality Eq.(\ref{duali}) holds.


\paragraph{Model.--} We consider $N$ particles pinned at random positions $\{r_n\}$ by an external 1D potential. Each particle has two
internal states, $\{ \uparrow, \downarrow \}$, which can be treated as (pseudo-)spin states. The particles experience
power-law spin exchange described by the XY Hamiltonian:
\begin{equation}
H_{XY}=-\frac{J}{2} \sum_{n,m} \frac{1}{|r_n-r_m|^a} \left ( S_n^+ S_m^- + S_n^- S_m^+ \right ),
\end{equation}
with $S^{\pm}_n$ the spin operators associated to particle $n$.
Assuming that all particles are in the state $\downarrow$, we denote $| n \rangle$ the state with a single spin-flipped particle $\uparrow$ at the site $n$ placed at position $r_n$.
The propagation of a single excitation among the particles, determined by the XY exchange, can be modeled by the Hamiltonian:
\begin{equation}
H=\sum_n \sum_{m\neq n} H_{n,m} |n\rangle\langle m|.
\label{eq:Model}
\end{equation}
The hopping of the excitation between the $n$--th and $m$--th particles follows a power law
\begin{equation}
H_{n,m}=- J |r_n-r_m|^{-a}.
\end{equation}
Below we assume for simplicity $J=1$. Since the positions $r_n$ are randomly distributed, the excitation experiences a peculiar randomness
that may significantly handicap its propagation in the system. The single-particle model is expected to describe well
the case of more than one excitation, as long as the gas of excitations is dilute enough.


\paragraph{Localization and multifractality--} We are interested in the localization properties of the
eigenstates $|\psi_s\rangle$ of $H$, $H |\psi_s\rangle = E_s |\psi_s \rangle$,
close to the maximum of the density of states (see Suppl. Material),
and, in particular, how these properties depend on $a$. The eigenfunctions and eigenenergies were obtained by exact diagonalization of
Model~\eqref{eq:Model} with $N$ up to $8.4\times 10^4$.

We characterize the eigenstates $|\psi_s \rangle=\sum_n \psi_s(n) |n\rangle$ by the moments $I_q(s)=\sum_n|\psi_s(n)|^{2q}\propto N^{-\tau(q)}$, where
$D(q)=\tau(q)/(q-1)$ are the so-called fractal dimensions. Localized states are characterized by $D(q)=0$, ergodic extended states by $D(q)=1$, while {\it extended, but non-ergodic, multifractal states} have a set of non-trivial fractal dimensions $0<D(q)<1$ \cite{Wegner1981,Altshuler1986,Evers2008,DeLuca2014,DengAltshuler2016}.
The Legendre transform $\tau(q)=q\alpha-f(\alpha)$ defines the multi-fractal spectrum~(MFS) $f(\alpha)$~\cite{Wegner1981}, which
characterizes the Hausdorff dimension of the manifold of sites where $|\psi_s(n)|^2=N^{-\alpha}$. Normalization, $\sum_{n}|\psi_{s}(n)|^2=1$, requires $\alpha\geq 0$.

An important difference between power-law- and short-range-hopping models is that in the former case localized states, if they exist at all, have power-law decaying
tails $|\psi_{s}(n)|^2\propto 1/|n-n_{0}|^{\gamma}$, i.e. they are {\it algebraically} localized.
One can show that the corresponding $f(\alpha)$ has a triangular form, where $f(\alpha)=k\alpha$ for $0<\alpha<1/k$ with
$k=d/\gamma$.
This is in contrast to {\it exponentially localized} states for a short-range hopping, where $k=0$. Note that $|\psi_{s}(n)|^{2}\propto |H_{n+n_{0},n_{0}}|^{2}\propto n^{-2a}$
in the {\it perturbative} regime at $a>d$, and hence $\gamma$ cannot be smaller than $\gamma_{a}=2a > 2d$. Therefore, for algebraically localized states in 1D one has $k<1/2$ like for the exponentially localized states on the Bethe lattice \cite{DeLuca2014}.

\begin{figure}[t]
\begin{center}
\includegraphics[width =1.0\columnwidth]{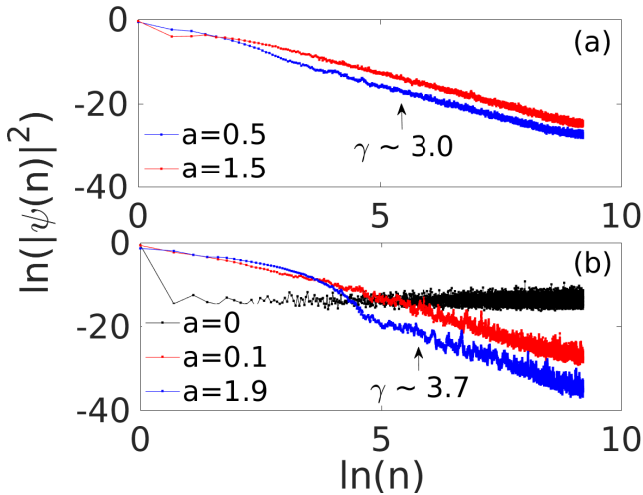}
\caption{Average of $\ln(|\psi(n)|^2)$ for 1D chain of $N=2\times 10^4$ sites vs. logarithm of the distance $n$ from the localization center. (a) PLE model with random site positions and deterministic hopping function $1/r^{a}$ for $a=0.5$ and $a=1.5$. The duality Eq.(\ref{duali}) manifests itself in the same slope $-\gamma(a)$ of the curves at the tail.
(b) Model of Ref.\cite{Celardo2016} on regular lattice  with diagonal disorder and deterministic power-law hopping for $a=0$, $a=0.1$ and $a=1.9$. Notice a drastic difference between the exactly solvable  \cite{Yuzbashyan} case $a=0$ and the two dual cases with $a$ close to zero and $2$ ($a=0.1$ and $a=1.9$). At $a\rightarrow 0$ the eigenstates are algebraically localized with  $\gamma\rightarrow 4$, while at $a=0$ they are "critically
multifractal" with $f(\alpha)=\alpha/2$ for $0<\alpha<2$ and  have a form of a sharp peak on the top of a background [23]. In contrast to the eigenstates of the standard Anderson model where the background is exponentially small, the critically multifractal states are characterized by much stronger background $\sim N^{-2}$.}
\label{fig:algebraic}
\end{center}
\end{figure}

We evaluate the spatial distribution of the eigenstates by setting the localization center placed at the maximum value of $|\psi|^2$, at index $n=n_{0}$, and
averaging $\ln |\psi(n)|^2$ over disorder realizations and over an energy window.
For the localized states discussed below the  typical average ${\rm exp}(\overline{\ln|\psi(n)|^2})$ is well fitted by the expected algebraic dependence $|n-n_{0}|^{-\gamma}$ as seen in Fig.\ref{fig:algebraic}, except for the special point $a=0$ (see Fig.2b) where the PLE model is disorder-free and the model [21] is exactly solvable [24].


\begin{figure*}[t]
\begin{center}
\includegraphics[width =2\columnwidth]{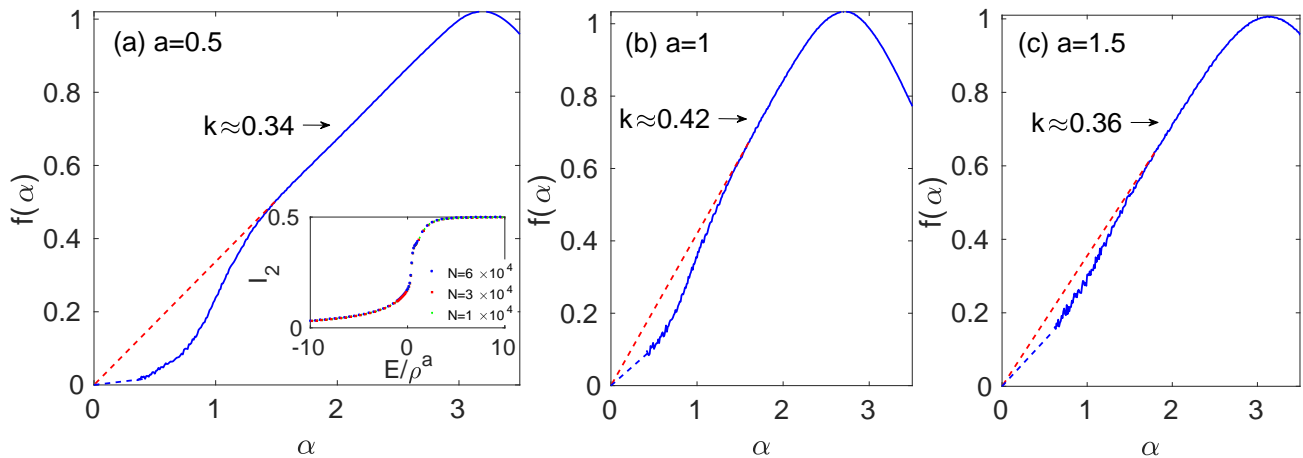}
\caption{Multi-fractal spectrum $f(\alpha)$ (blue curve) obtained from the distribution function $P(|\psi_{s}(n)|^{2})$, extrapolated to $N\to\infty$, for (a) $a=0.5$; (b) $a=1$; (c) $a=1.5$. We consider eigenstates at the maximum of the density of states. The inset of (a) shows that the inverse participation ratio $I_2$ does not depend on $N$ thus proving insulating behavior. Note that the concave feature in $f(\alpha)$ appears in all cases (see text). The red dashed line in each case shows the triangular   $f_{tr}(\alpha)$ with $k<1/2$ that corresponds to the same moments $I_q$ as $f(\alpha)$. The blue dashed line indicates that $f(0)=0$, which is confirmed by our numerics.}
\label{fig:MFS}
\end{center}
\end{figure*}



\paragraph{Multi-fractal spectrum.--} We obtain the MFS in the thermodynamic limit by means of a linear extrapolation of our finite-size calculations in terms of $1/\ln N$~\cite{DeLuca2014,DengAltshuler2016}.
For $a>1$ we expect localized states, and indeed we find linear $f(\alpha)$ with a slope $k< 1/2$.
At $a=1$, for all eigenstates the MFS displays a triangular shape with the slope $k_c\simeq 0.42$~(Fig.~\ref{fig:MFS}(b)).
Since $k_c$ is very close to the critical value $1/2$, the eigenstates for $a=1$ are either weakly localized or critical.
For $a<1$  one would naively expect extended eigenstates. However,  this expectation appears to be wrong.
The eigenstates inside a band emerging due to disorder-broadening of the macroscopically degenerate level {\it remain localized} for all $0<a<1$.

Note that     the slope $k$ of the linear part of $f(\alpha)$ shown in Fig.~\ref{fig:MFS}  reflects the algebraic decay of $|\psi_s(n)|$ far from the localization center. For small $\alpha$ the MFS becomes {\it concave}, implying a reduced probability to find intermediate values $|\psi_s(n)|^2\propto N^{-\alpha}$. A similar concave MFS has been discussed for the insulating phase of the Rosenzweig-Porter random matrix ensemble~\cite{KravtsovNJP2016}. The moments $I_q\propto N^{-\tau(q)}$ are however determined by finding the tangent $-\tau(q)+ q\, \alpha$ to $f(\alpha)$ that does not cross the MFS at any point. Hence, $\tau(q)$ and fractal dimensions for $f(\alpha)$ with a concave segment are the same as those for the MFS in which the concave part is replaced by the red dashed line shown in Fig~\ref{fig:MFS}.


\paragraph{Duality.--}
The localization properties of the eigenstates of Model~\eqref{eq:Model} for $0<a<1$ and for $1<a<2$ exhibit a striking {\it duality} given by Eq.(\ref{duali}).
The slope $k$ of the MFS is depicted in the upper panel of Fig.~\ref{Fig:duality}. The lower panel
shows the exponent $\gamma$ extracted from the average $\overline{\ln(|\psi(n)|^2)}\propto -\gamma\ln{|n-n_{0}|}$~(see Fig.~\ref{fig:algebraic}).
Both graphs are in good mutual agreement given by the relation $k=1/\gamma$.
Remarkably, a similar duality is present for the model of Ref.~\cite{Celardo2016}, where disorder is purely diagonal~(see Fig.\ref{Fig:duality}).


\paragraph{Proof of duality for $a\rightarrow 0$ and $a\rightarrow 2$.--}
As was already mentioned, for $a=0$ there is an $(N-1)$-fold degeneracy of the excited states, which may be chosen both extended and localized on equal footing.
The localization properties of degenerate states are controlled by the perturbation $V(r)=r^{-a} -1$;
namely, by the properties of the  "good zero-order wave functions"~\cite{Landau-Lifshitz},
which are eigenvectors of the perturbation operator in a certain convenient basis.
Let us select a basis of $(N-1)$  states including $N/2$ non-overlapping dimers,
$|\psi_{n,n+1}\rangle=(|n\rangle-|n+1\rangle)/\sqrt{2}$, $N/4$ non-overlapping  tetramers, $|\psi_{n,n+1}^{n+2,n+3}\rangle=(|n\rangle+|n+1\rangle-|n+2\rangle-|n+3\rangle)/2$,
$N/8$ octamers, and so on.  The basis constructed in this way is orthonormal and complete for any $N=2^k$, where $k$ is a positive integer.
Moreover, the $n$-mers are compact and thus represent localized states for any $n={\cal O}(1)$.
The perturbation results in hopping between $n$-mers.
The hopping amplitude $V_{n,m}$ from the dimer $|\psi_{n,n+1}\rangle$ to the dimer $|\psi_{m,m+1}\rangle$, located far away~(and $m\gg n$), in the {\it limit of small} $a$ is given by:
\begin{eqnarray}
V_{n,m}&=&\Big\{\frac{1}{|r_{m+1}-r_{n}|^{a}}-\frac{1}{|r_m-r_n|^{a}} \nonumber \\
&+&\frac{1}{|r_m-r_{n+1}|^{a}} -\frac{1}{|r_{m+1}-r_{n+1}|^{a}}\Big\}\propto \frac{a}{R_{nm}^{2}},
\end{eqnarray}
where $R_{nm}=(r_{m+1}+r_m-r_{n+1}-r_n)/2$. A similar dependence holds for the hopping between tetramers and all $n$-mers~(see Suppl. Mat.).

Therefore "good zero-order wave functions"  in the basis of $n$-mers  obey the same equation as eigenvectors of the Hamiltonian with  hopping $\sim 1/r^{2}$.  Thus, we have reduced the problem of localization close to the point of macroscopic degeneracy $a=0$ to the problem  of localization in a system of $n$-mers with hopping amplitude $\propto 1/R_{nm}^{2}$ for which localization is well understood.
This proves the localization in our model  for $a\rightarrow 0$ case \cite{footnote-a0} and its duality to localization at $a=2$.

Note finally that for $a<1$ at the edge of the spectrum at negative energy there are also delocalized states~(see Suppl. Material). The measure of the number of such states is zero, but they will be responsible for the transport in the system.


\paragraph{Role of sign-alteration.--} The localization properties of our model and, in particular, the duality (2) are drastically different from those of
PLBRM \cite{MirlinFyod1996} and Levitov's scenario~\cite{Levitov1990}. We show at this point that this difference is related to the sign-randomness in the long-range hopping. To this end we modify our model in such a way that $H_{n,m}$  acquires a randomness $\eta_{n,m}$:
\begin{equation}
H_{n,m}=-(J+\eta_{n,m})/ r_{nm}^a,
\label{eq:modified}
\end{equation}
where $\eta_{n,m}$ are random bounded numbers, $|\eta_{n,m}|<W$, and $r_{nm}=|r_{n}-r_{m}|$. Model~\eqref{eq:Model} corresponds to $W=0$, $J=1$, while
that of Ref.~\cite{MirlinFyod1996} corresponds to $W=1$, $J=0$.
Model~\eqref{eq:modified} can be physically realized for a spatially pinned two-component system, in which the spin-like excitations may be transferred either between particles of the same or of different species (see Suppl. Material).

Another possible modification is the staggering model:
\begin{equation}\label{staggering}
H_{n,m}=(-1)^{r_{nm}}/ r_{nm}^a,
\end{equation}
for which the sign of the product $H_{m,j_{1}}H_{j_{1},j_{2}}...H_{j_{\ell-1},n}$ over a path of $\ell$ hops from $m$ to $n$
is independent of the number of hops and equal to $(-1)^{r_{nm}}$.

The analysis presented in Suppl. Material shows that even at small $W=0.1$, Model~\eqref{eq:modified} exhibits for $J=1$ the same properties
as PLBRM and the eigenstates are extended for $a<1$. In contrast, Model~\eqref{staggering} is in the same universality class as Model~\eqref{eq:Model}.
This shows that localization in Model~\eqref{eq:Model} is due to {\it interference of long-range paths} involving one or more hops,
which is not affected by staggering, but destroyed by sign-randomness.


\paragraph{Conclusions.--} Systems with power-law, $\propto 1/r^a$, spin-exchange between randomly spaced particles represent a peculiar form of off-diagonal disorder that leads to
surprising localization properties. We have shown that 1D spin excitations remain algebraically localized
for any value of the hopping power $a> 0$. Moreover, we show that our model is a representative of a new universality class of models with power-law hopping,
characterized by a duality Eq.(\ref{duali}) between models with $0<a<1$ and $2>a>1$.  


We acknowledge fruitful discussions with B.L. Altshuler, I.Khaymovich and E. Yuzbashyan. X. D. and L. S. thank the support of the DFG (SFB 1227 DQ-mat and FOR2247). G. V. S. acknowledges funding from the European Research Council under European Community's Seventh Framework Programme (FP7/2007-2013 Grant Agreement no. 341197).



\begin{thebibliography}{99}

\bibitem{DePaz2013} A.~de~Paz, A.~Sharma, A.~Chotia, E.~Mar\'echal, J.~ H.~Huckans, P.~Pedri, L.~Santos, O.~Gorceix, L.~Vernac, and B.~Laburthe-Tolra, Phys. Rev. Lett. {\bf 111}, 185305 (2013).

\bibitem{Yan2013} B.~Yan, S.~A.~Moses, B.~Gadway, J.~P.~Covey, K.~R.~A.~Hazzard, A.~M.~Rey, D.~S.~Jin, and J.~Ye, Nature {\bf 501}, 521 (2013).

\bibitem{Saffman2010} M.~Saffman, T.~G.~Walker, and K.~M\o lmer, Rev. Mod. Phys. {\bf 82}, 2313 (2010).

\bibitem{Waldherr2014} G.~Waldherr, Y.~Wang, S.~Zaiser, M.~Jamali, T.~Schulte-Herbrueggen, H.~Abe, T.~Ohshima, J.~Isoya, P.~Neumann, and J.~Wrachtrup, Nature {\bf 506}, 204 (2014).

\bibitem{Alvarez2015} G.~A.~\'Alvarez, D.~Suter, and R.~Kaiser, Science {\bf 349}, 846 (2016).

\bibitem{Richerme2014} P.~Richerme, Z.-X.~Gong, A.~Lee, C.~Senko, J.~Smith, M.~Foss-Feig, S.~Michalakis, A.~V.~Gorshkov, and C.~Monroe, Nature {\bf 511}, 198 (2014).

\bibitem{Jurcevic2014} P.~Jurcevic, B.~P.~Lanyon, P.~Hauke, C.~Hempel, P.~Zoller, R.~Blatt, and C.~F.~Roos, Nature {\bf 511}, 202 (2014).

\bibitem{Hung2016} C.-L.~Hung, A.~Gonzalez-Tudelac, J.~I.~Cirac, and H.~J.~Kimble, PNAS {\bf 113} E4946 (2016).

\bibitem{Levitov1990} L.~S.~Levitov, Phys. Rev. Lett. {\bf 64}, 547 (1990).

\bibitem{AltshAleiEfet2011}I.~L.~Aleiner, B.~L.~Altshuler, and K.~B.~Efetov, Phys. Rev. Lett. {\bf 107}, 076401 (2011).

\bibitem{MirlinFyod1996} A.~D.~Mirlin, Y.~V.~Fyodorov, F.~M.~Dittes, J.~Quezada, and T.~H.~Seligman, Phys. Rev. E {\bf 54}, 3221 (1996).

\bibitem{KrMut1997} V.~E.~Kravtsov and K.~A.~Muttalib, Phys. Rev. Lett. {\bf 79}, 1913 (1997).

\bibitem{KrYev2003} O.~Yevtushenko and V.~E.~Kravtsov, J. Phys. A-Math. Gen. {\bf 30},
8265(2003).

\bibitem{YevOss2007} O.~Yevtushenko and A.~Ossipov, J. Phys. A-Math. Gen. {\bf 40},
4691(2007).

\bibitem {Cuevas2007} E.~Cuevas and V.~E.~Kravtsov, Phys. Rev. B {\bf 76}, 235119 (2007).

\bibitem{Chalk2010} V.~E.~Kravtsov, A.~Ossipov, O.~M.~Yevtushenko, and E.~Cuevas, Phys. Rev. B {\bf 82}, 161102(R) (2010).

\bibitem{Evers2008} A.~D.~Mirlin, Y.~V.~Fyodorov, A.~Mildenberger, and F.~Evers, Phys. Rev. Lett. {\bf 97}, 046803 (2006).

\bibitem{RMTbook} V.~E.~Kravtsov, in: Handbook on random matrix theory
(Oxford University Press, 2010), arXiv:0911.0615v1.

\bibitem{Bogomol2003} E.~Bogomolny, O.~Bohigas and C. ~ Schmit, J. Phys. A: Math. Gen. {\bf 36}, 3595 (2003).

\bibitem{Ossipov2012} A.~ Ossipov, J. Phys. A: Math. Gen.  {\bf 46}, 105001  (2013).

\bibitem{Celardo2016} G.~L.~Celardo, R.~Kaiser, and F.~Borgonovi, Phys. Rev. B {\bf 94}, 144206 (2016).

\bibitem{footnote-a0} Note that the limit $a\rightarrow 0$ does not commute with the limit $N\rightarrow\infty$ of infinite system size.  The duality holds only if the limit $N\rightarrow\infty$ is taken first. The opposite order of limits corresponds to the exceptional point $a=0$ where the PLE model is disorder-free and macroscopically degenerate, whereas  for the model \cite{Celardo2016} with diagonal disorder   the wave functions in the bulk of the spectrum are critically multifractal. In this case the problem  allows for an exact solution \cite{Yuzbashyan}.

\bibitem{Celardo2013} A. Biella, F. Borgonovi, R. Kaiser, G.L.
Celardo, EuroPhys. Lett. 103, 57009 (2013).

\bibitem{Yuzbashyan} J. A. Scaramazza, B. S. Shastry, E. A. Yuzbashyan, Phys. Rev. E {\bf 94}, 032106 (2016).

\bibitem{MezarParisiZee1999} M.~Mezard, G.~Parisi, and A.~Zee, Nucl. Phys. B {\bf 559},
689 (1999).

\bibitem{Dyson1953}F.~J.~Dyson, Phys. Rev. {\bf 92}, 1331 (1953).

\bibitem{Lifshitz1964} I. M. Lifshitz, Adv. Phys. {\bf 13}, 483 (1964); Soviet Physics Uspekhi {\bf 7}, 549 (1965).

\bibitem{Brouwer2000} P.~W.~Brouwer, C.~Mudry, A.~Furusaki, Phys. Rev. Lett {\bf 84}, 2913 (2000).

\bibitem{Wegner1981} F.~Wegner, Z. Phys. B {\bf 44}, 9 (1981).

\bibitem{Altshuler1986} B. L. Altshuler, V. E. Kravtsov, and I. V. Lerner, JETP Lett. {\bf 43}, 441 (1986);
 B. L. Altshuler,  V. E. Kravtsov, and I. V. Lerner, Zh. Eksp. Teor. Fiz. {\bf 91}, 2276 (1986)~(Sov. Phys. JETP {\bf 64}, 1352 (1986)).

\bibitem{DeLuca2014} A.~De~Luca, B.~L.~Altshuler, V.~E.~Kravtsov, and A.~Scardicchio, Phys. Rev. Lett. {\bf 113}, 046806 (2014).

\bibitem{DengAltshuler2016} X.~Deng, B.~L.~Altshuler, G.~V.~Shlyapnikov, and L.~Santos,
  Phys. Rev. Lett. {\bf 117}, 020401 (2016).

\bibitem{KravtsovNJP2016} V.~E.~Kravtsov, I.~M.~Khaymovich, E.~Cuevas, and M.~Amini,
New Journal of Physics {\bf 17}, 122002 (2015).

\bibitem{Landau-Lifshitz} L. D. Landau and E. M. Lifshitz, {\it Quantum mechanics}, Pergamon Press, 1991.

\end{thebibliography}
\end{document}


\title{Duality in power-law localization in disordered one-dimensional systems:\\
Supplemental Material}
\author{X. Deng, V.E. Kravtsov, G.V. Shlyapnikov, and L. Santos}

\date{\today}

\maketitle

\renewcommand\thefigure{S\arabic{figure}}

\section{Density of states}
\begin{figure}[t]
\begin{center}
\includegraphics[width=0.5\columnwidth]{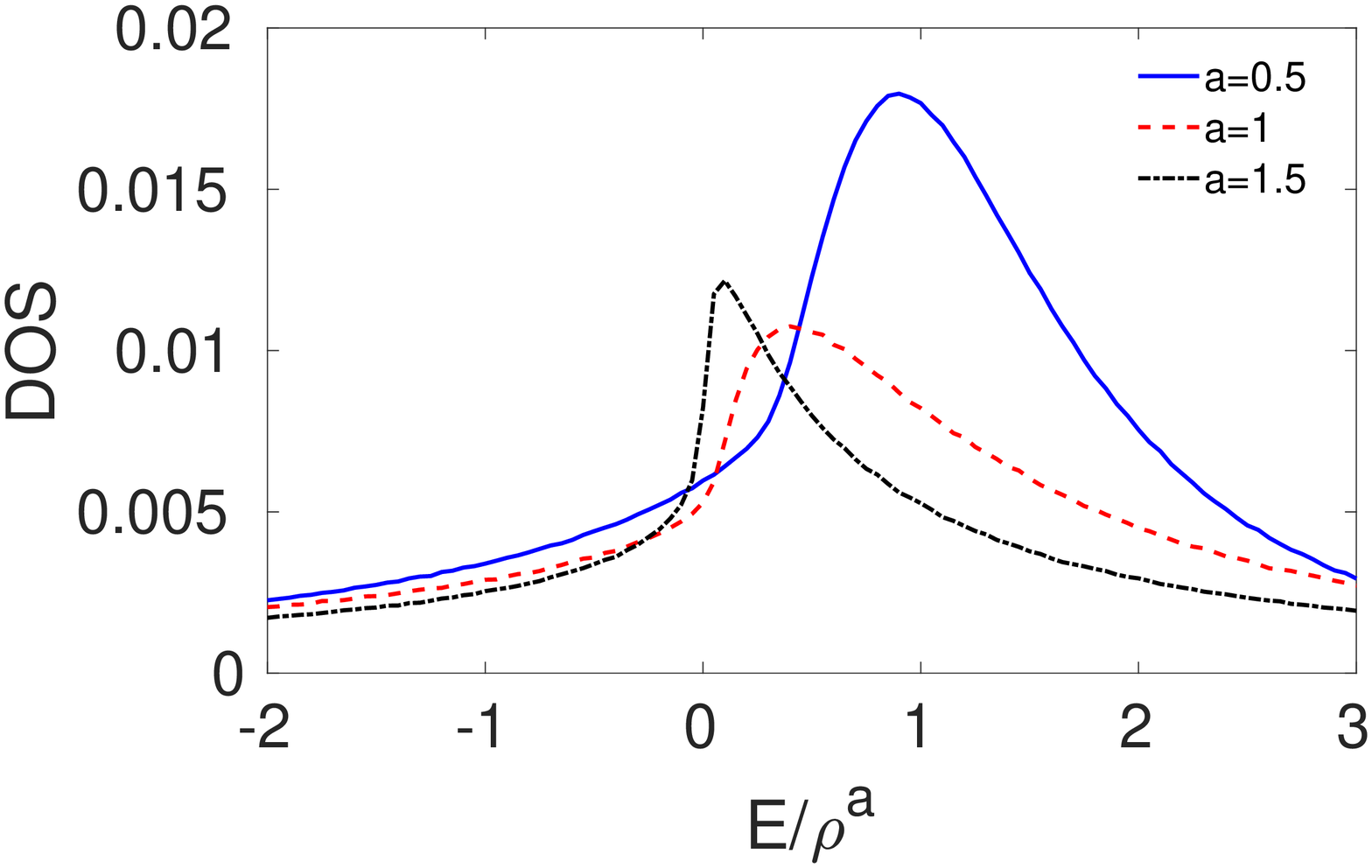}
\caption{Density of states for models with power-law exponents $a=0.5$, $a=1$, and $a=1.5$. The energy has been re-scaled as $E/\rho^a$ to properly compare the different cases.}
\label{fig:S0}
\end{center}
\end{figure}

The localization properties of the system are best characterized by analyzing the eigenstates
at the maximum of the density of states~(DOS). As shown in Fig.~\ref{fig:S0}, for Model~(3) of the main text the DOS is not symmetric with a maximum at $E=0$, as it is e.g. for the PLBRM model.
This is best understood by considering an infinitely long-range power law, $a=0$. In this case, the DOS is basically a singular peak at $E=1$, since
$N-1$ states are degenerate with energy $E=1$~\cite{Ossipov2012, Celardo2016}. For $a>0$, the DOS broadens and its maximum is shifted towards $E=0$. For $a\gg 1$, the DOS approaches
a peak at $E=0$~\cite{footnote-S1} but in contrast to the nearest-neighbor hopping models with off-diagonal disorder, where the chiral symmetry enforce the symmetric form of DOS, it remains highly asymmetric. Our results on $f(\alpha)$ and $\overline{|\psi(n)|^2}$ shown in the main text and below are evaluated at the DOS maximum, but we obtain similar results in a broader range of energies.

\section{Hopping of tetramers and higher $n$-mers}
For a total particle number $N$ that is not equal to $2^k$, with $k$ being a positive integer, one should also include trimers, sectamers, etc. in the basis set of the wave functions.
The $n$-mer states with $n$ close to $N$ are already not compact. However, the number of such extended states has measure zero, and they can be omitted. The derivation of the hopping amplitude for
$n$-mers with a large $n$ is quite cumbersome, and for this reason we only present it here for tetramers. The related amplitude is given by:
\begin{eqnarray}
&&1/r_{n,m}^a  +1/r_{n,m+1}^a - 1/r_{n,m+2}^a  -1/r_{n,m+3}^a +1/r_{n+1,m}^a + 1/r_{n+1,m+1}^a - 1/r_{n+1,m+2}^a  -1/r_{n+1,m+3}^a \nonumber \\
&-&1/r_{n+2,m}^a  -1/r_{n+2,m+1}^a + 1/r_{n+2,m+2}^a +1/r_{n+2,m+3}^a -1/r_{n+3,m}^a -1/r_{n+3,m+1}^a + 1/r_{n+3,m+2}^a  +1/r_{n+3,m+3}^a\nonumber  \\
&=&4/r_{nm}^{a}  +1/r_{n,m+1}^a  + 1/r_{n+1,m}^a -1/r_{n,m+3}^a -1/r_{n+3,m}^a    -2/r_{n+2,m}^a      - 2/r_{n,m+2}^a \approx -16a(1+a) / R_{nm}^{2+a}.
\end{eqnarray}

\section{Staggering and randomization of sign of hopping amplitude}

\begin{figure}[t]
\begin{center}
\includegraphics[width =0.5\columnwidth]{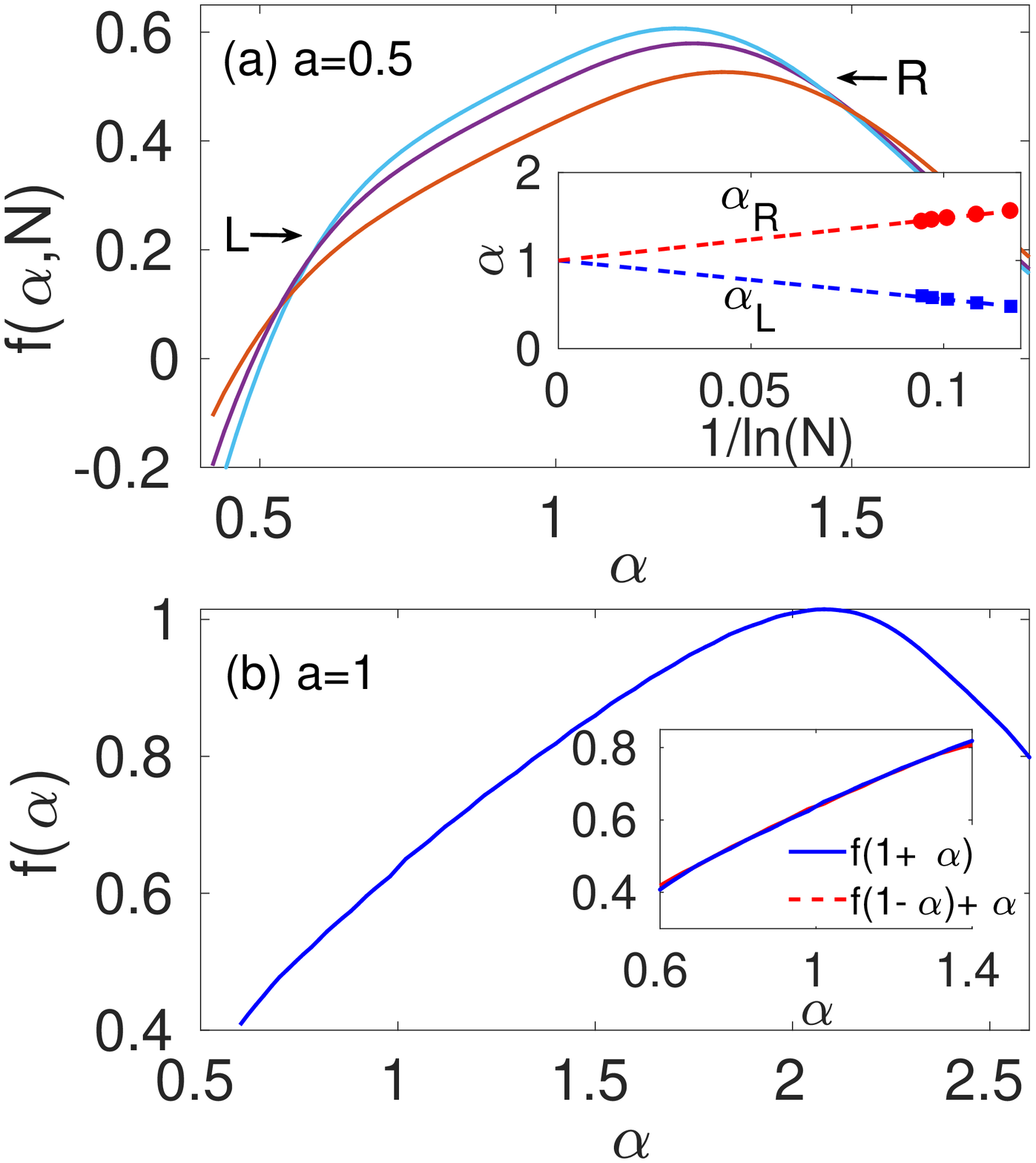}
\caption{Modified model. MFS for Model (8) for (a) $a=0.5, E=0$ and (b) $a=1, E/\rho=0.5$, where $W=0.1$, $J=1$. In (a) the finite-size $f(\alpha,N)$ cross each other to the left ($L$) and to the right ($R$) of $\alpha=1$. From bottom to top $N=10^4, 3\times 10^4, 6\times 10^4$. The extrapolation of the crossings ($\alpha_L$ and $\alpha_R$) for $1/\ln(N)\rightarrow 0$ linearly approaches $1$ (see the inset in (a)). This supports the ergodic extended states. In (b) $f(\alpha)$ presents the parabolic dependence characteristic of non-ergodic extended states. They have the symmetry property $f(1+\alpha)=f(1-\alpha)+\alpha$ \cite{Evers2008}, which is confirmed in the inset.
}
\label{fig:S1}
\end{center}
\end{figure}
Consider Model~(6) of the main text, where the sign of the single-excitation hopping acquires a random part.
As already mentioned in the main text, Model~(6) can be physically realized considering a spatially pinned two-component system, in which the spin-like excitations may be transferred either between particles of the same or of different species. Due to the different intra- and inter-species spin exchanges, the excitation hop is different depending on the components involved. If the spatial pinning is species-insensitive,
for any given realization of the pinning potential there will be a random distribution of the two components. As a result, a spin-like excitation transferred among the particles
will experience not only the power-law decay, but also an additional random prefactor depending on the components between which the hop occurs. This corresponds to Model~(6).
In this case, the $|n-m|^{-2a}$ contributions of $\langle H_{n,m}^2\rangle$ and $\langle H_{n,m}\rangle^2$ do not cancel each other, and hence $\Delta H_{n,m} \sim |n-m|^{-2a}$ for $|n-m|\gg 1$.
In this sense, Model~(6) has the same localization properties as the standard PLBRM model~(despite
of not being exactly identical).

Our numerical results show that $a\leq 1$ is an extended case~(see Fig.~\ref{fig:S1}). Indeed, the two intersection points in $f(\alpha)$ at $a=0.5$ for different system sizes are both extrapolated to 1, just confirming a collapse of $f(\alpha)$ into a vertical line at $\alpha=1$, which corresponds to the ergodic case. In contrast, for $a=1$ Model~(6) shows a clear multifractal behavior that obeys the Mirlin-Fyodorov symmetry $f(1+\alpha)=f(1-\alpha)+\alpha$.

In the picture of hopping of $n$-mers the replacement
$$1/r_{nm}^{a}\rightarrow (1+\eta_{nm})/r_{nm}^{a}$$
destroys the cancelation of $1/R^{1+a}$ terms in the hopping amplitudes. So, in the zero-order approximation the system is critical and not localized like without the $\eta_{nm}$ term. 
This opens the possibility for the system to be delocalized at a finite $a<1$.

Consider now Model~(7) of the main text, i.e. the staggered model, in which $1/r_{nm}^{a}$ is replaced as 
$$1/r_{nm}^{a}\rightarrow (-1)^{r_{nm}}/r_{nm}^{a}.$$
Such a replacement does not spoil the cancelation of $1/R^{1+a}$ terms both in the hopping amplitudes of dimers and those of tetramers, etc. However, it
leads to the cancelation of the $1/R^{2+a}$ terms in the hopping amplitudes of tetramers and higher $n$-mers. This means that they will be less involved in the zero-order approximation than in the case of non-staggered hopping. At the same time, no such cancelation happens in the hopping matrix elements of dimers. Thus, we arrive at the conclusion that staggering does not weaken localization at small $a$.

\begin{figure}[t]
\begin{center}
\includegraphics[width =0.5\columnwidth]{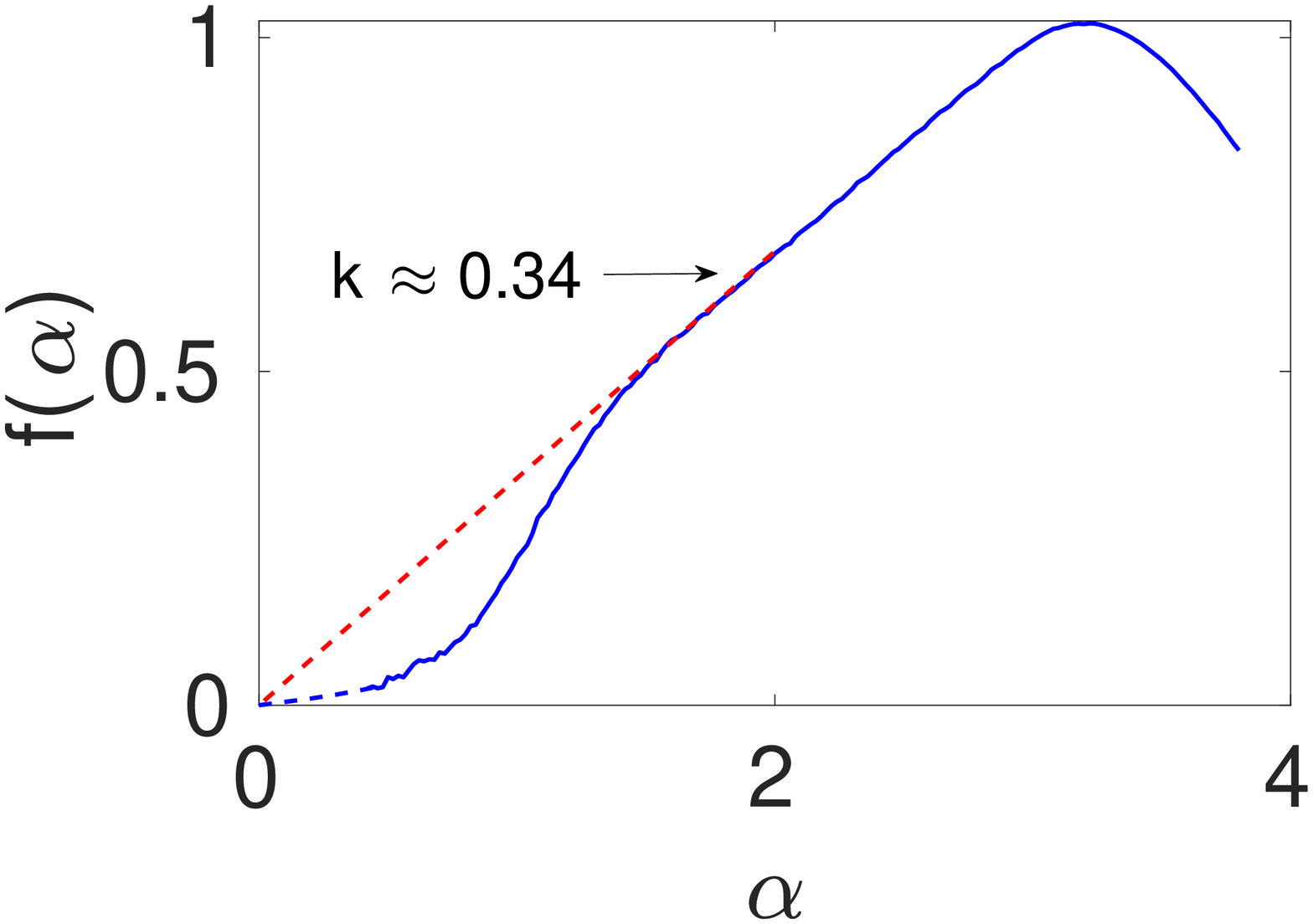}
\caption{ $f(\alpha)$ for a staggered model at $a=0.5$. Also for the staggered mode the states are localized.}
\label{fig:stagg}
\end{center}
\end{figure}

Indeed, the analysis of $f(\alpha)$ for this model shows~(see Fig.~\ref{fig:stagg}) that the distribution of wave function amplitudes for the staggered model is practically indistinguishable from that of the model without staggering. This indicates that tetramers and higher $n$-mers play only a small role in the $n$-mer picture of localization.

\section{Delocalized states at $a<1$}

\begin{figure}[t]
\begin{center}
\includegraphics[width =0.5\columnwidth]{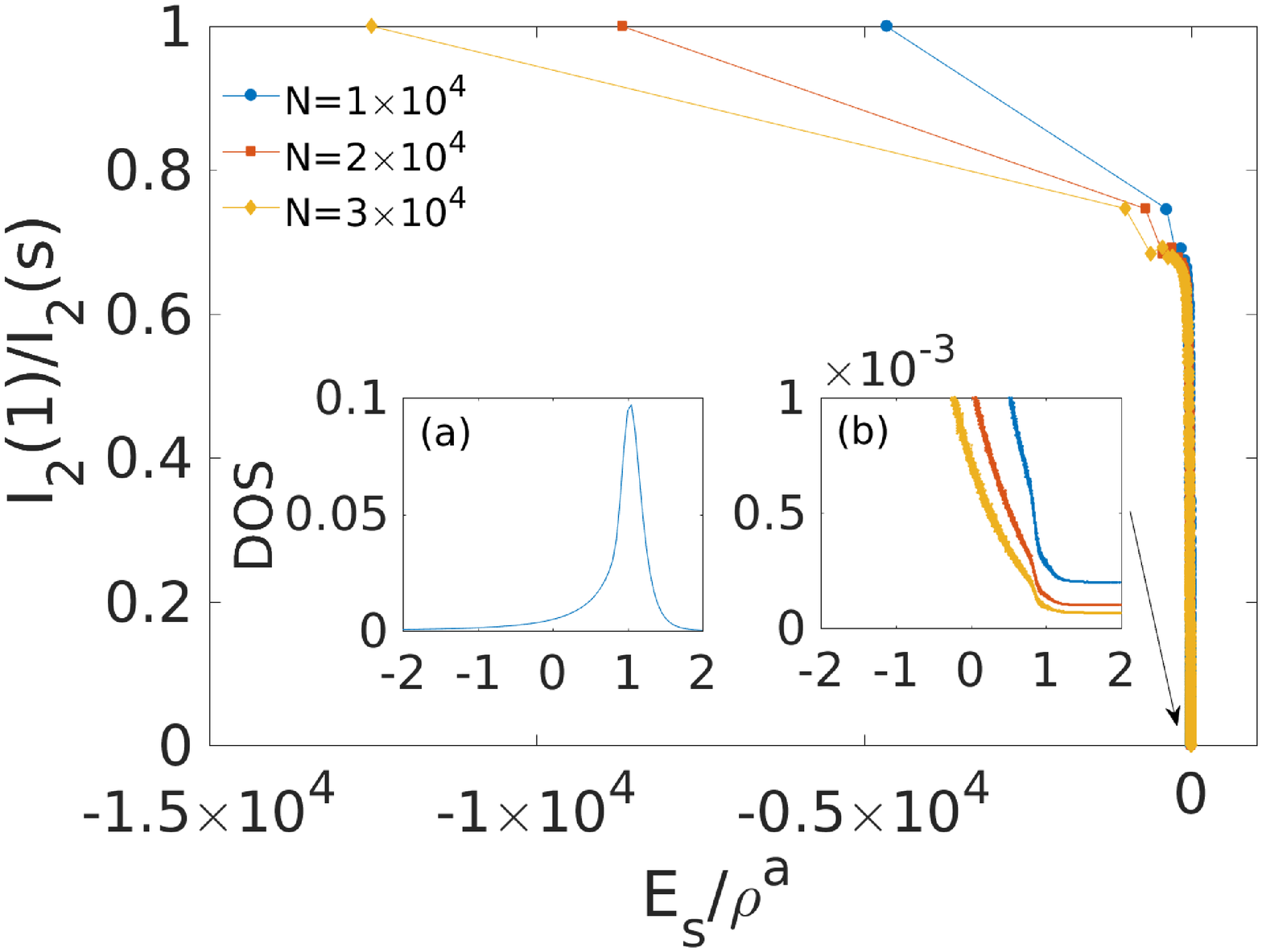}
\caption{The ratio $I_2(1)/I_2(s)$ for $a=0.1$, and $N=1\times10^4$, $2\times10^4$, and $3\times10^4$. Inset (a) depicts the DOS, whereas inset (b) shows an enlarged view of $I_2(1)/I_2(s)$ at $E \approx 0$.}
\label{fig:S3}
\end{center}
\end{figure}

As mentioned in the main text, we also have delocalized states at $a<1$, although the number of such states has measure zero for $N$ tending to infinity. This is already seen in the theoretical limit $a\rightarrow 0$, for which there 
is at least one delocalized state, namely the state with energy $-(N-1)$~\cite{Ossipov2012, Celardo2016}. For finite $a$ the hybridization leads to the appearance of several other extended states. This is clearly seen from the calculation of the inverse participation ratio $I_2(s)$ for eigenstates with energy $E_s$. In Fig.~\ref{fig:S3} we show the quantity $I_2(1)/I_2(s)$ as a function of $E_s$ for $a=0.1$, where $I_2(1)$ is the inverse participation ratio for the delocalized state with the lowest energy. There are several delocalized states at the lowest eigenenergies, and then at higher energies the ratio $I_2(1)/I_2(s)$ sharply drops to zero indicating that the states are localized.

\begin{figure}[t]
\begin{center}
\includegraphics[width =0.5\columnwidth]{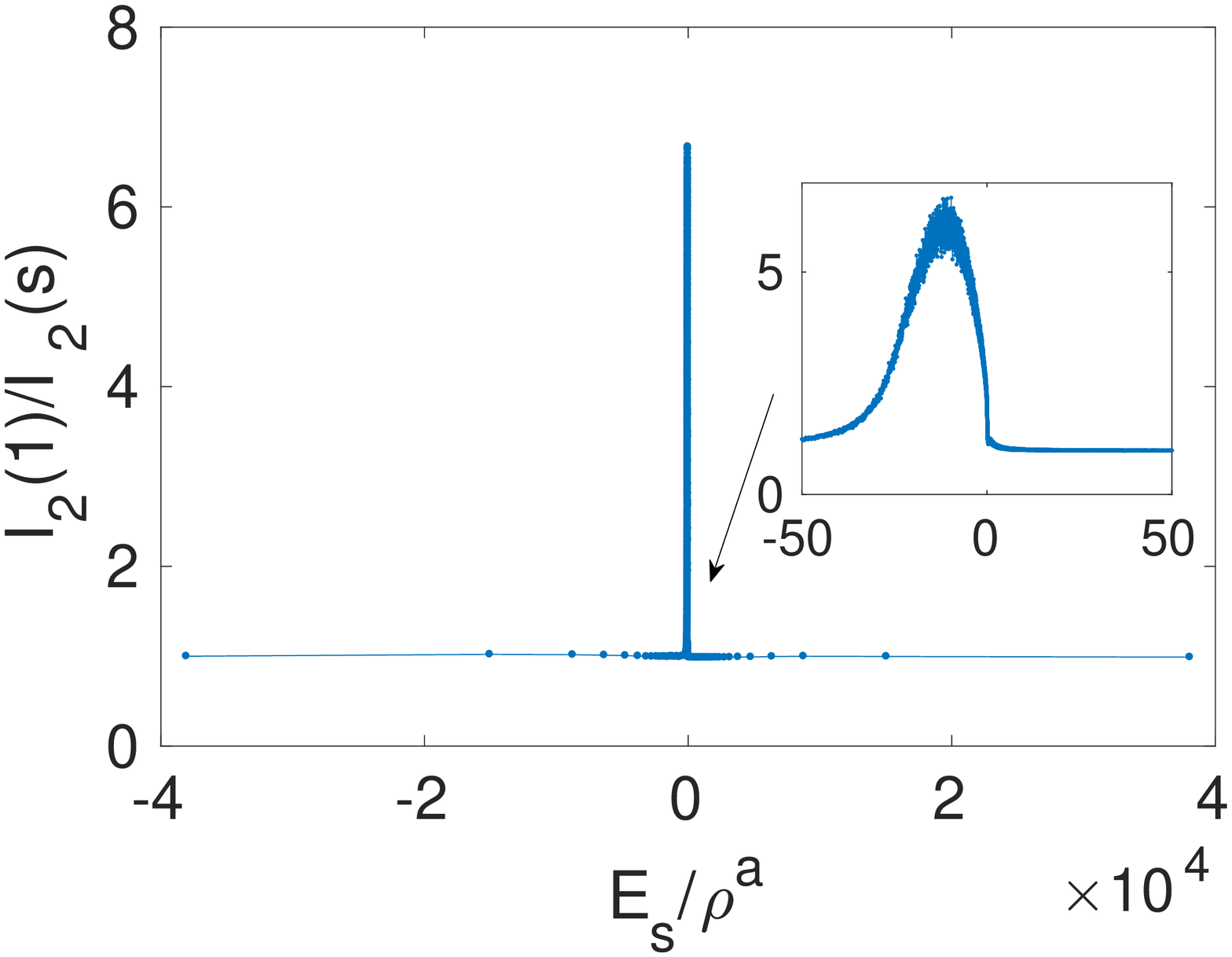}
\caption{The ratio $I_2(1)/I_2(s)$ for $a=1$ and $N=2\times 10^4$. The inset shows an enlarged view of the peak appearing at $E\approx0$.}
\label{fig:S4}
\end{center}
\end{figure}

Delocalized states disappear at $a\geq 1$\cite{footnote-S2}. In Fig.~\ref{fig:S4} we show the quantity $I_2(1)/I_2(s)$ for $a=1$. In this case the lowest energy state is localized, as well as all other states. The only feature to mention is that close to $E=0$
the quantity $I_2(1)/I_2(s)$ sharply increases, which indicates that the states have a much larger localization radius. This is expected since for $a\gg 1$ the problem is close to that with nearest neighbor hopping,
where all states are localized but the localization radius diverges when approaching $E=0$ \cite{Dyson1953,Brouwer2000}.

